\documentstyle[aps,epsf,multicol]{revtex}

\def\ro{\rho}
\def\be{\begin{equation}}
\def\ee{\end{equation}}

\def\etal{{\it et al.~}}

\begin{document}
\draft
\title{Collective motion of self-propelled particles: kinetic phase transition
in one dimension}
\author{Andr\'as Czir\'ok$^{1}$\thanks{
E-mail address: {\tt czirok@hercules.elte.hu}},
Albert-L\'aszl\'o Barab\'asi$^2$ and Tam\'as Vicsek$^{1}$
}

\address{
$^1$  Department of Atomic Physics,
E\"otv\"os University, Budapest, Puskin u. 5-7, 1088 Hungary \\
$^2$ Department of Physics, University of Notre Dame, Notre Dame, IN 46556\\
}

\date{\today}
\maketitle

\begin{abstract}
We demonstrate that a system of self-propelled particles (SPP) exhibits
spontaneous symmetry breaking and self-organization in one dimension,
in contrast with previous analytical predictions. To explain this
surprising result we derive a new continuum theory that can account for
the development of the symmetry broken state and belongs to the same 
universality class as the discrete SPP model.
\end{abstract}


\begin{multicols}{2}

The transport
properties of systems consisting of self-propelled particles (SPP) have
generated much attention lately \cite{VCBCS95,CV95,TT,jpn,albano,Bus}.
This interest has been largely
motivated by analogous processes taking place in numerous biological phenomena
(e.g., bacterial migration on
surfaces \cite{bac}, flocking of birds, fish,
quadrupeds\cite{bio}, correlated motion of ants\cite{Millonas} and 
pedestrians\cite{helb0})
as well as in various other systems, including
driven granular materials \cite{DHT,Hem95} and traffic models \cite{Nagel}. 
The models describing these phenomena are distinctively non-equilibrium, 
exhibiting kinetic phase
transitions and self-organization, of particular interest from the point of
view of modern statistical mechanics\cite{evans}.

In the simplest version of the SPP model \cite{VCBCS95} -- introduced
to study collective biological motion --
 each particle's velocity is set to a fixed magnitude, $v_0$.
 The
interaction with the neighboring particles changes only the direction of motion:
the particles tend to
align their orientation to the local average velocity.
Numerical simulations in 2D provided evidence
of a second order phase transition \cite{CSV}
between an ordered phase in which the mean velocity of the entire system,
 $\langle v \rangle$, is nonzero and a disordered phase with
$\langle v \rangle = 0$, as
the strength of the noise is increased or the density of the particles
decreased. 

This SPP model is similar to the $XY$ model of classical
magnetic spins because the velocity of the particles, like the
local spin of the $XY$ model, has fixed length and continuous rotational
symmetry. In the $v_0=0$ and low noise limit the
model reduces {\it exactly} to a Monte-Carlo dynamics of the $XY$ model.
Since the $XY$ model does {\it not} exhibit a long-range ordered phase at
temperatures $T>0$\cite{KT}, the ordered state observed in \cite{VCBCS95} is
surprising.
To explain this discrepancy, Toner and Tu (TT) \cite{TT} proposed
a continuum theory that included in a self-consistent way the
non-equilibrium effects as well. They have shown that their model is
different from the XY model for $d<4$ and found an ordered phase in $d=2$
\cite{racz}.
While TT could account for the first
time for the ordered phase in 2D SPP models, their theory does not allow 
for an ordered phase for $d=1$.

In this paper we demonstrate that kinetic phase transition
and ordering takes place in 1D as well. This result, not foreseen by
the existing analytical approaches, motivated us to introduce a new
continuum theory describing the SPP model in arbitrary dimensions. 
Linear stability analysis shows that indeed in 1D the continuum
model exhibits ordering. Furthermore, numerical investigations
indicate that the continuum theory and the discrete SPP model
belong to the same universality class.

{\it The 1D SPP model ---} Let us consider $N$ off-lattice particles along a
line of length $L$. The particles are characterized by their coordinate
$x_i$ and dimensionless velocity $u_i$ updated as
\begin{eqnarray}
x_i(t+1) = x_i(t) + v_0 u_i(t), \nonumber\\
u_i(t+1) = G\Bigl(\langle u(t) \rangle_i\Bigr) + \xi_i.
\label{EOMD}
\end{eqnarray}
The local average velocity $\langle u \rangle_i$ for the $i$th particle is
calculated over the particles located in the interval
$[x_i-\Delta,x_i+\Delta]$, where we fix $\Delta=1$.  The function $G$
incorporates both the propulsion and friction forces which set the velocity
in average to a prescribed value $v_0$:
$G(u)>u$ for $u<1$ and $G(u)<u$ for $u>1$ \cite{noteI}.
The distribution function $P(x=\xi)$ of the noise $\xi_i$ is uniform
in the interval $[-\eta/2,\eta/2]$.

Keeping $v_0$ constant ($v_0 = 0.1$), the adjustable control
parameters of the model are the average density of the particles,
$\rho=N/L$, and the noise amplitude $\eta$.
We implemented one of the
simplest choices \cite{noteII} for $G$,
\begin{equation}
G(u)=\cases{
	(u+1)/2 & for $ u > 0$ \cr
	(u-1)/2 & for $ u < 0$, \cr
     }
\end{equation}
and applied random initial and periodic boundary conditions.

In Fig. 1 we show the time evolution of the model for 
$\eta=2.0$. In a short time the system reaches 
an ordered state, characterized by a spontaneous broken symmetry and
clustering of the particles. In contrast, for larger values of $\eta$ a 
disordered velocity field can be found.

{\it Scaling and exponents ---}
To capture quantitatively the transition from
an ordered to a disordered state, in Fig. 2a we plot the order parameter
$\phi\equiv\langle u \rangle$ vs $\eta$ for
various $\rho$.
As in the two dimensions \cite{VCBCS95,CSV}, 
the ordered phase emerges through a second order phase transition.
Near the critical noise amplitude, $\eta_c(\rho,L)$, which
separates the ordered from the disordered phase,
$\phi$ vanishes as
\be
\phi(\eta,\rho)\sim \cases{
         \Bigl({\eta_c(\ro,L) - \eta\over \eta_c(\ro,L)}\Bigr)^\beta
                & for $\eta<\eta_c(\ro,L)$ \cr
        0  & for $\eta>\eta_c(\ro,L)$ \cr
    },
\label{scale}
\ee
which finding is supported by the increasing scaling
regime with the system size $L$ (Fig. \ref{f1}b) and by the
convergence of $\eta_c({\rho},L)$ to a non-zero
$\eta_c(\rho,\infty)$ value.
We find that $\beta=0.60\pm0.05$, which is different
from both the mean-field value $1/2$ \cite{HESBOOK} and $\beta=0.42\pm0.03$
found in $d=2$ \cite{CSV}.

Fig. 2a also shows that the various $\phi(\eta,\rho)$ curves can be collapsed onto
a single function $\phi_0(x)$, where $x=\eta/\eta_c(\rho)$,
just like in $d=2$. As shown in \cite{CSV},
the consequence of this fact is that near the critical density the order
parameter vanishes as
\be
\phi(\eta,\rho)\sim \cases{
         \Bigl({\rho - \rho_c(\eta,L)\over \ro_c(\eta,L)}\Bigr)^{\beta'}
                & for $\ro>\ro_c(\eta,L)$ \cr
        0  & for $\ro<\ro_c(\eta,L)$ \cr
    },
\ee
with $\beta'=\beta$.
These results can be summarized in the $\ro-\eta$ phase diagram shown in
Fig. 2c. We also find that the critical line, $\eta_c(\rho)$, follows
\begin{equation}
\eta_c(\rho)\sim\rho^\kappa,
\label{kappa}
\ee
with $\kappa=0.25\pm0.05$.

While the above numerical results demonstrate the existence of the phase
transition in one dimension and provide numerical values for the scaling
exponents $\beta$, $\beta'$ and $\kappa$, the origin of these values is
unclear at this point. In particular, the emergence of the ordered phase 
in 1D is not predicted either by the equilibrium theories or by the TT  model.
To overcome this
limitation next we introduce and investigate a set of continuum
equations
(which can be generalized to any dimension) in terms of $U(x,t)$
and $\varrho(x,t)$, where $U$ and $\varrho$ represent the coarse-grained
dimensionless velocity and density fields, respectively.  Analogous 
approaches were fruitful in the study of a similar SPP system, one-lane traffic 
flow\cite{Helbing}.

{\it Continuum theory ---}
Let us denote by $n(u,x,t)dudx$ the number of particles moving with a velocity
in the range of $[v_0u,v_0(u+du)]$ at time $t$ in the $[x,x+dx]$ interval.
The particle density $\varrho(x,t)$ is then given as $\varrho=\int ndu$,
while the local dimensionless average velocity $U(x,t)$ can be calculated
as $\varrho U=\int nudu$.
According to the microscopic
rules of the dynamics, in a given time interval $[t,t+\tau]$ all particles
choose a certain velocity $v/v_0=[G(\langle u \rangle) + \xi]$ and travel
a distance $v\tau$. Thus, the time development of the ensemble average
(denoted by overline) of $n$ is governed by the master equation
$\overline{n(u,x,t+\tau)}=
\overline{\varrho(x',t)}p(u\vert U(x',t))$, where $x'=x-v_0u\tau$ and
 $p(u\vert U)$
denotes the conditional probability of finding a particle with a
velocity $u$ when the local velocity field $U$ is given. From Eq. (\ref{EOMD})
we have $p(u\vert U)=P(u-G(\langle U \rangle))$.
Since $n$ is finite, the actual occupation numbers in a given system differ
from $\overline{n}$. This fact can be accounted by adding an intrinsic
noise term to the master equation as
\be
n(u,x,t+\tau) = \varrho(x',t)p(u\vert U(x',t) ) +
\nu(u,x',t),
\label{boltz}
\ee
where $\nu$ has the following properties:
(i) $\overline{\nu}=0$, (ii) due to the conservation of the particles
$\int\nu du=0$, and (iii) since 
we have a random sampling process, the actual values of $n$ satisfy Poisson
statistics, i.e., the distribution function of $\nu$ depends on $\varrho$, 
$u$ and $U$ as 
$P(\nu)=\lambda^{\nu+\lambda}\exp(-\lambda)/\Gamma(\nu+\lambda+1)$, where
$\lambda=\varrho p(u|U)$. Thus, we have $\overline{\nu^2}=\overline{n}$.

Taking the Taylor expansion of $n(u,x-v_0u\tau,t)$
up to second order in $x$ and integrating
Eq. (\ref{boltz}) according to $du$, in the $v_0\tau\ll1$,
$\sigma^2\equiv\int P(u)u^2du\gg1$, $\varrho\gg1$ and
$v_0\tau\sigma^2<1$ limit we obtain
\be
\partial_t\varrho = -v_0 \partial_x(\varrho u) +
D\partial_x^2\varrho,
\label{CEOM1}
\ee
where $D=v_0^2\tau\sigma^2/2$.
Note, that the appearance of the diffusion
term is a consequence of the
non-vanishing correlation time $\tau$.
Since $\int p(u\vert U)udu = G(\langle U \rangle)$, integrating
Eq.(\ref{boltz}) according to $udu$, expanding $\langle U \rangle$
as $\langle U \rangle = U + [\partial_x^2 U + 2(\partial_x U)
(\partial_x \varrho)/\varrho]/6$ \cite{CsCz}, using
(\ref{CEOM1}), we arrive at
\be
\partial_tU=f(U)+\mu^2\partial_x^2 U +
\alpha{(\partial_x U)(\partial_x \varrho)\over\varrho} + \zeta,
\label{CEOM2}
\ee
where $f(U)=(G(U)-U)/\tau$, $\mu^2=(dG/dU)/(6\tau)$, $\alpha=2\mu^2$ and
$\zeta=\int \nu udu/\varrho\tau$. Note, that $f(U)$ is an antisymmetric
function with $f(U)>0$ for $0<U<1$ and $f(U)<0$ for $U>1$,
$\overline{\zeta}=0$, and $\overline{\zeta^2}=\sigma^2/\varrho\tau^2$.

At this point we consider Eqs. (\ref{CEOM1}) and (\ref{CEOM2}) with
the coefficiens $\mu$, $\alpha$, $\sigma$, $v_0$ and $D$ as
the continuum theory describing a large class of SPP models. These
equations differ from both the equilibrium field theories
and the nonequilibrium system investigated by TT \cite{TT}. 
The main differences are due to 
(i) the nonlinear coupling term $(\partial_x U)(\partial_x \varrho)/\varrho$,
and (ii) the statistical properties of the noise $\zeta$. 
For $\alpha=0$ the dynamics of
the velocity field $U$ is independent of $\varrho$, and Eq.(\ref{CEOM2})
is equivalent
to the $\Phi^4$ model describing spin chains, where
domains of opposite magnetization develop at finite temperatures
\cite{HESBOOK}.

To study the effect of the nonlinear term in (\ref{CEOM2}),
we now investigate the development of the ordered phase in
the deterministic case ($\sigma=0$).
For $\alpha=0$ Eqs. (\ref{CEOM1}) and (\ref{CEOM2}) have a set
of (meta)stable stationary solutions
$\varrho^*$ and $U^*$ describing a ``domain wall'' separating
two regions with opposite velocities. 
Since we can freely translate these solutions, 
we assume $U^*(0)=0$.
Performing linear stability analysis we next show that for certain finite
values of $\alpha$ the above stationary solutions are unstable.

{\it Linear Stability Analysis ---}
In the following we make use of the fact that
the dynamics of $\varrho$ is very slow compared to that of $U$.
Let us write $U$ in the form of $U(x,t)=U_0(x,t)+u(x,t)$,
where $U_0(x,t)=U^*(x-\xi(t))$ and $\xi(t)$ is defined by $U(\xi(t),t)=0$,
i.e. $\xi(t)$ defines the position of the domain wall.
Moreover, we substitute $f=U-U^3$ as one of the simplest
choices for $f(U)$.
Now in the $u\ll U_0$ and $\partial_xu\ll\partial_xU_0$ limit
in the moving frame $x'=x-\xi(t)$ Eq.(\ref{CEOM2}) reads as
\be
\partial_tu' = \dot{\xi}a + (g-2)u' + \mu^2\partial_x^2u' +
\alpha h(x'+\xi),
\label{LIN1}
\ee
where $u'(x')=u(x)$,
$[\partial_x\varrho^*(x)][\partial_xU_0(x)]/\varrho^*(x)=h(x'+\xi)$,
$\partial_xU^*(x')=a(x')$
and $g(x')=(df/dU)(U^*(x'))+2$. 
From $U(\xi(t),t)=U_0(\xi(t),t)=0$ we get
$u'(x'=0,t)=0$, which yields
\be
-\dot{\xi}a(0)=\mu^2\partial_x^2u'(0,t) + \alpha h(\xi).
\label{LIN2}
\ee
Now (\ref{LIN1}) and (\ref{LIN2}) describe the time development of
the velocity field in terms of $u'$ and $\xi$.

Fourier transforming (\ref{LIN1}) and (\ref{LIN2}) we find that the
short wavelength fluctuations ($k\rightarrow\infty$) are stabilized by the
Laplacian term in (\ref{LIN1}). However, the growth rate $\lambda$ of the long
wavelength fluctuations, defined by $\xi\sim\exp(\lambda t)$, is determined
by the characteristic equation
\be
(-2+4\mu^2-\lambda)({\alpha v_0\over\sqrt2D\mu}-\lambda) -
\alpha{10\sqrt{2}v_0\mu\over D} = 0.
\label{disp}
\ee
 It can be seen that
(\ref{disp}) has positive solution for large enough $\alpha$.
This result means that for $\alpha>\alpha_c$ 
($\alpha_c\approx2\sqrt2D\mu/v_0$)
the domain wall solution $U^*$ is 
unstable, hence in this regime the walls disappear and all particles
move in the same direction, {\it demonstrating that (\ref{CEOM1}) and (\ref{CEOM2})
predict an ordered phase in one dimension}.

To further confirm the relevance of the continuum theory to the discrete model (1),
we integrate numerically (\ref{CEOM1}) and (\ref{CEOM2}). The results of the
integration are in excellent agreement with the results obtained for the 
discrete
model and with the linear stability analysis, and can be summarized as follows:
(i) For the noiseless case ($\sigma=0$) we find that for $\alpha>\alpha_c$
there is an ordered phase, which disappear for $\alpha<\alpha_c$; (ii)
The ordered phase predicted by (\ref{disp}) is present
for the noisy $\sigma>0$ as well; (iii) Increasing $\sigma$ leads to a 
second order phase transition from the ordered to the disordered state.
Since $\sigma$ plays the role of $\eta$ in (\ref{EOMD}), this transition is
equivalent with (3) observed in the discrete model; (iv) Finally, we measured
the order parameter $\phi$ as a function of $\sigma$ for various values of $D$.
As Fig 2d. illustrates, for small $D$ Eq.(3) with $\beta=0.6$ provides an excellent 
fit to the numerical results, indicating that the discrete model (1) and the continuum
theory (\ref{CEOM1}) and (\ref{CEOM2}) {\it belong to the same 
universality class}.

In conclusion, we showed that the SPP model exhibits spontaneous
symmetry breaking and ordering in one dimension, a result
surprising both in the light of the equilibrium spin models, and
of the continuum theory investigated by TT. 
We introduced a new continuum
theory, whose terms are explicitelly derived from the ingredients of the discrete model
(in contrast with constructing it from symmetry arguments).
Linear stability analysis indicates that an ordered phase
can develop as the domain walls become unstable. 
 While here we limited ourselves to the 1D case of the
 continuum theory, (\ref{CEOM1}) and (\ref{CEOM2}) can be generalized
 to arbitrary dimensions, thus apply to the physically and biologically
 relevant two and three dimensional models as well.
 Since the continuum model investigated by us contains the major
ingredients of the SPP models, such as self-propulsion and local
reorientation of the velocity, we expect our results to apply to a wide
variety of systems made of SPP particles \cite{VCBCS95,CV95,albano,bac,bio}.

AC is grateful to H.E. Stanley for the hospitality during his visit
at the Center for Polymer Studies, Boston University, and to the
A Peregrinatio Foundation of the E\"otv\"os University.
We thank Z. R\'acz for his useful comments on the manuscript.
This work was supported by OTKA F019299 and FKFP 0203/1997. ALB was partially
supported by the NSF CAREER award DM297-01998 and by the donors of the
Petroleum Research Fund, administrated by the ACS.

\end{multicols}

\begin{figure}
\caption{The dynamics of the 1D SPP model for $L=300$, $\eta=2.0$ and $N=600$.
The darker gray level represents higher particle density. Note that the particles 
exhibit clustering and the spontaneous broken symmetry of motion.}
\label{f0}
\end{figure}

\begin{figure}
\caption{a: The order parameter $\phi$ vs the noise amplitude normalized by
the critical amplitude $\eta_c(\rho)$, for $L=1000$ and various values of 
$\rho$. For $\eta<\eta_c(\rho)$ the system is in a symmetry-broken state
indicated by $\phi > 0$.  b: $\phi$ vanishes as a power-law
in the vicinity of $\eta_c(\rho)$.
Note the increasing scaling regime with increasing $L$. 
The solid line is a power-law fit with an exponent $\beta=0.6$, while the 
dotted line shows the mean-field slope $\beta=1/2$ as a comparison.
c: Phase diagram in the $\rho-\eta$ plane. The critical line follows
$\eta_c(\rho)\sim\rho^\kappa$. The solid curve represents a fit with 
$\kappa=1/4$.
d: The order parameter vs the std deviation of the noise
normalized by $\sigma^*=\sigma_c(D=0)$, obtained by direct numerical integration
of the continuum model for $\alpha=2$, $\mu=1$, $v_0=0.1$, $\rho=1$, 
$L=1000$ and various values of $D$. For $D\ll1$ $\phi(\sigma)$ follows
a power-law with an exponent $\beta=0.6$ (solid line).}
\label{f1}
\end{figure}


\begin{references}
\bibitem{noteI} In the 1D case we cannot apply the constant
velocity constraint of \cite{VCBCS95} as it strongly discretizes
the system leading to a discrete state cellular automaton.
\bibitem{noteII} The results presented here do not depend on the properties of
$G(v)$ at $v\approx 0$. We performed simulations with a continuous
choice of $G(v)$ as well, and the scaling properties remained unchanged.
\end{references}
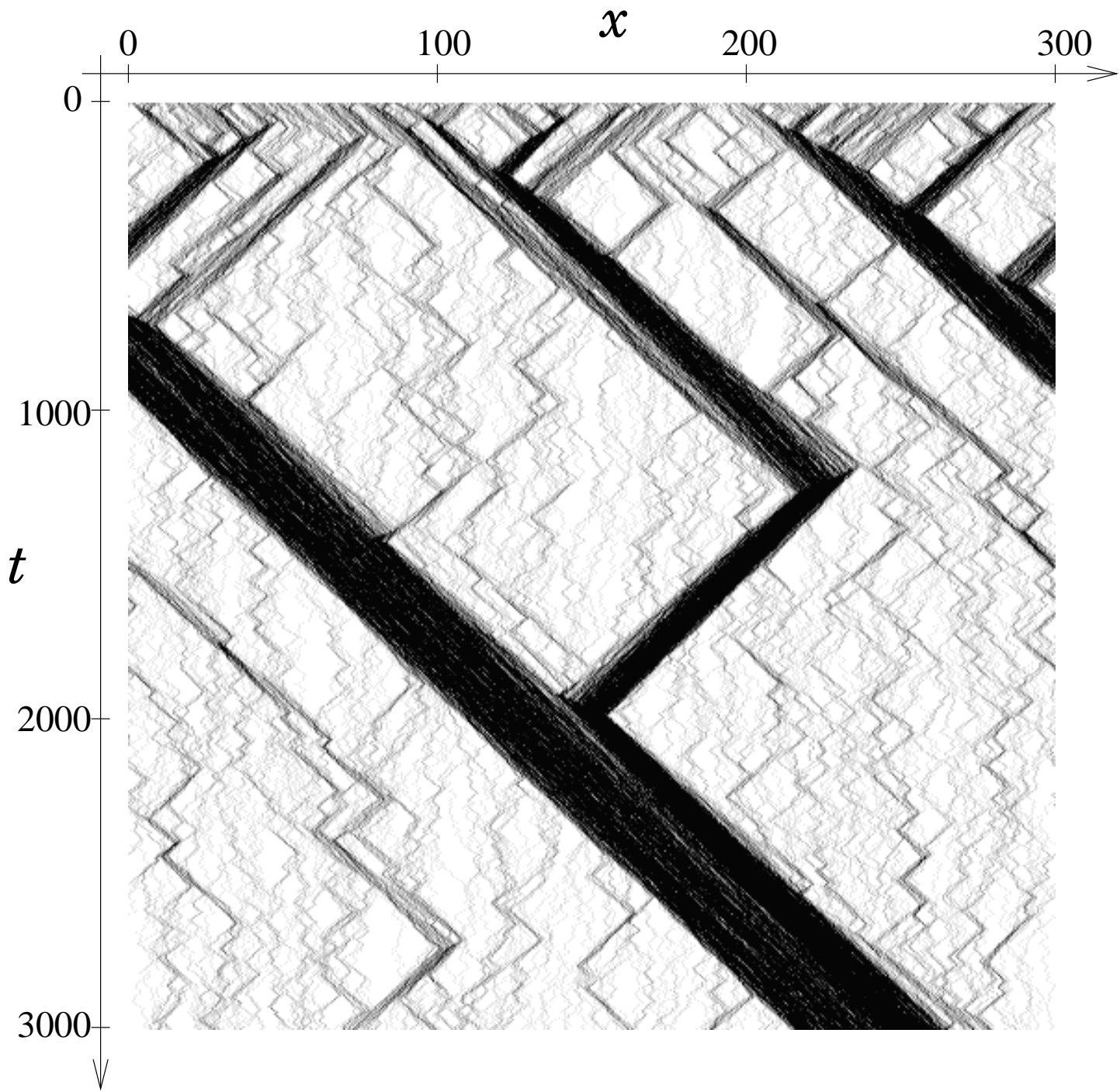
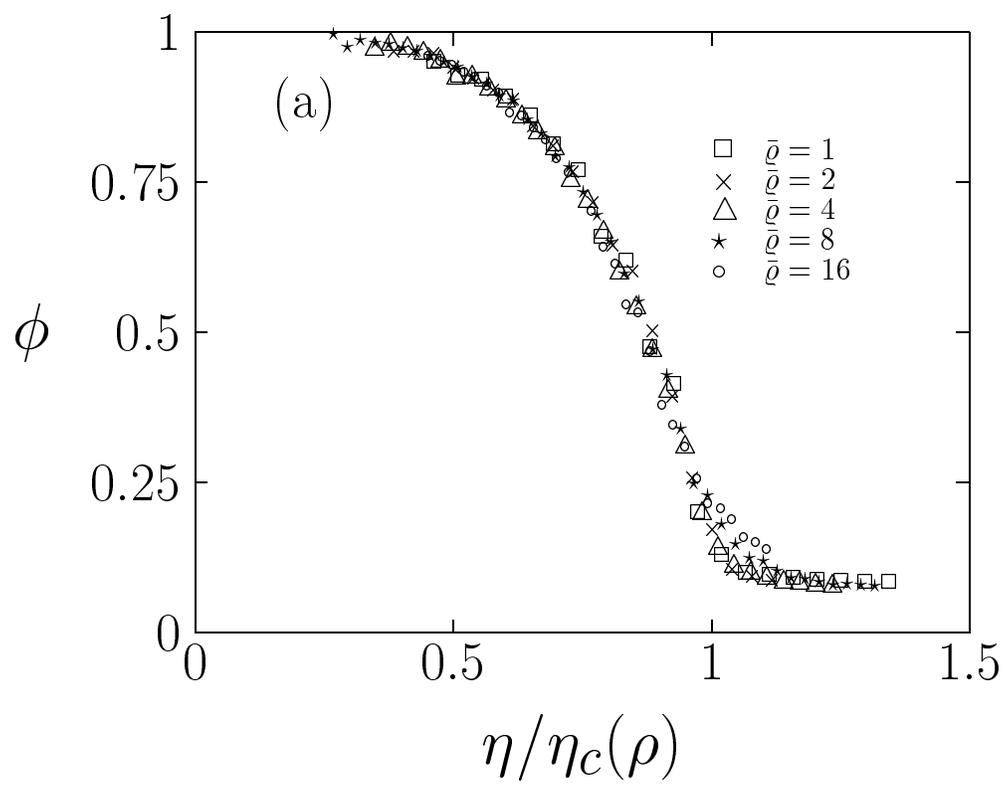
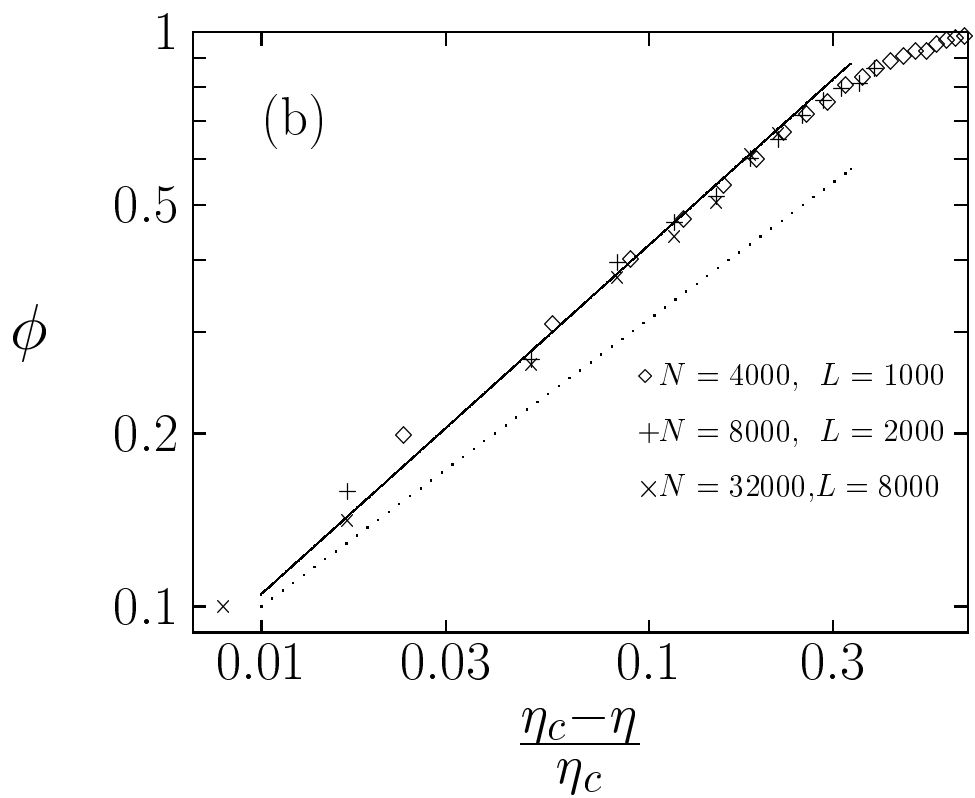
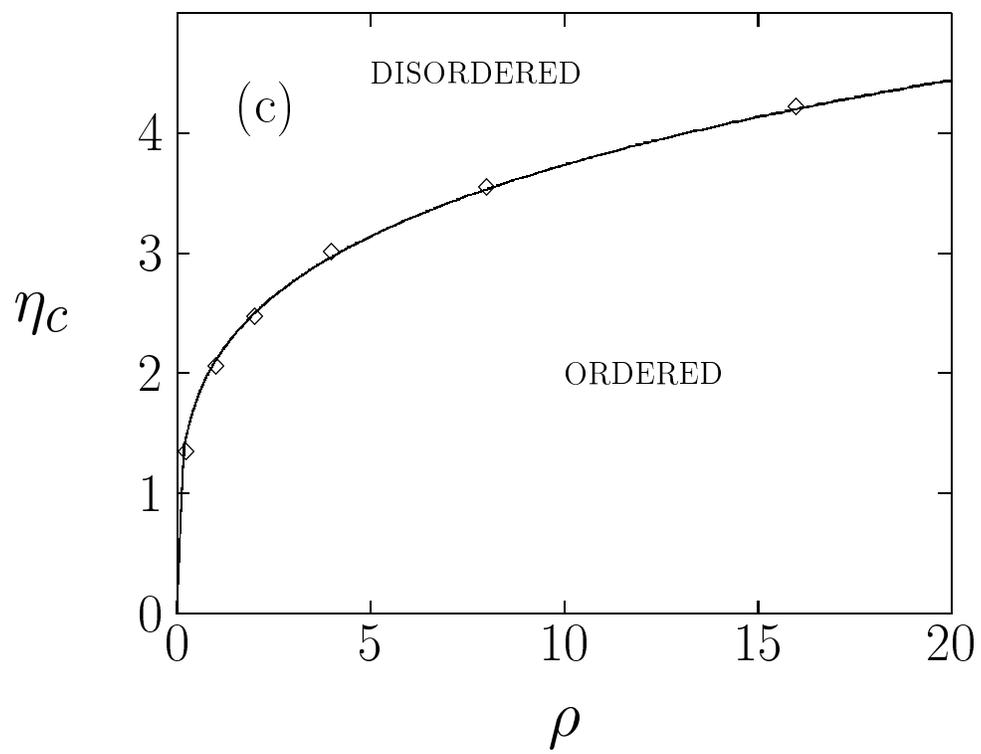
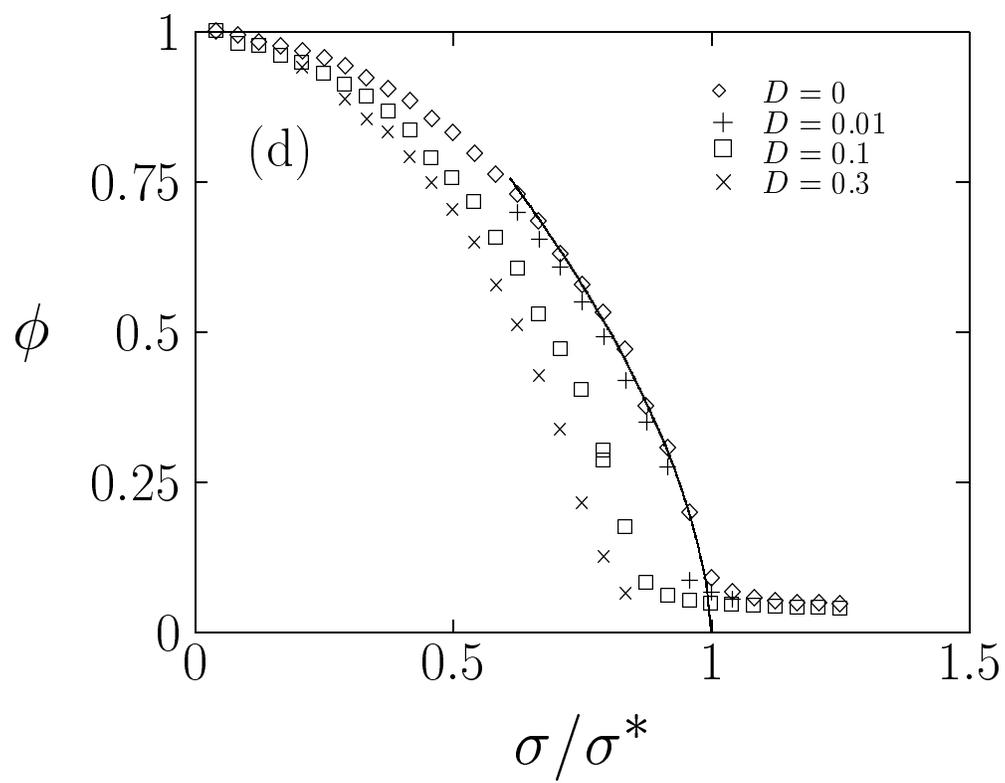
\end{document}